\documentstyle[pra,aps,twocolumn,epsf]{revtex}

\begin{document}
\draft
\twocolumn[\hsize\textwidth\columnwidth\hsize\csname @twocolumnfalse\endcsname

\title{
Conceptual design of nanostructures for efficient photoinduced phase transitions}

\author{
Tohru Kawamoto\footnote{E-mail: tohru.kawamoto@aist.go.jp} and Shuji Abe\footnote{E-mail: s.abe@aist.go.jp}
}

\address{
Nanotechnology Research Institute (NRI) and Research Consortium for Synthetic Nano-Function Materials Project (SYNAF),  
National Institute of Advanced Industrial Science and Technology (AIST), 1-1-1 Umezono, Tsukuba 305-8568, Japan
}
\date{\today}
\maketitle

%
%

\begin{abstract}
By means of Monte Carlo simulations on a kinetic model, we demonstrate that the efficiency of a photoinduced phase change can in general be enhanced drastically by using a superstructure of an appropriate combination of two components. 
This is due to the accelerated nucleation of converted domains in the structural blocks relatively close to local instability. 
The present mechanism provides a general guideline on the design of photocontrollable materials with potential applications for memory and storage devices.
\end{abstract}
\vskip2pc]


\narrowtext

Control of material phases by external stimuli, especially by light, has been a subject of much interest in recent years in view of potential applications for memory and storage devices in the future. 
It requires a material exhibiting bistability between two distinct phases. 
In many cases the {\it global} bistability of a solid has a {\it local} origin\cite{Sato96Science2,Hauser92CPL,Ogawa00PRL,Koshihara99JPCB}, while the photo-conversion between phases usually proceeds in a {\it cooperative} fashion\cite{Ogawa00PRL,Koshihara99JPCB}. 
This phenomenon is therefore called a photoinduced phase transition, in analogy with ordinary (thermally induced) phase transitions.

The actual observation of the phenomenon has been limited within a small number of materials: one or a few examples from each class of materials, including  spin-crossover complexes\cite{Ogawa00PRL}, Prussian blue analogues\cite{Sato96Science2,Sato97JES}, organic charge-transfer complexes\cite{Koshihara99JPCB} and conjugated polymers\cite{Koshihara92PRL}.   
The scarcity of materials is presumably related to the requirement that the two phases must be energetically close to each other, otherwise the metastable phase would return to the stable phase quickly. 

Here we propose a general scheme to overcome this difficulty.
Suppose we have a bistable material $\alpha$ in which the metastable state B is much higher in energy than the stable state A, so that photo-conversion is difficult. 
Let us then prepare a companion material $\beta$ in which the relative stability is opposite (see Fig.1(a)). 
If we combine the two materials in an appropriate fashion, e.g., in a superlattice as shown in Fig.1(b), and if there are sufficient cooperative interactions that favor the $\alpha$ and $\beta$ units being in the same state (either A or B), then the energy of the A phase in the total system can be made degenerate with that of the B phase. 
This is rather an obvious way of designing new suitable materials in general. 
However, this approach yields more than that: 
We will demonstrate in the following that the photo-conversion {\it efficiently is drastically enhanced} in such mixed structures compared with uniform structures. 

\begin{figure}
\begin{center}
  \leavevmode
  \epsfxsize=85mm
  \epsfbox{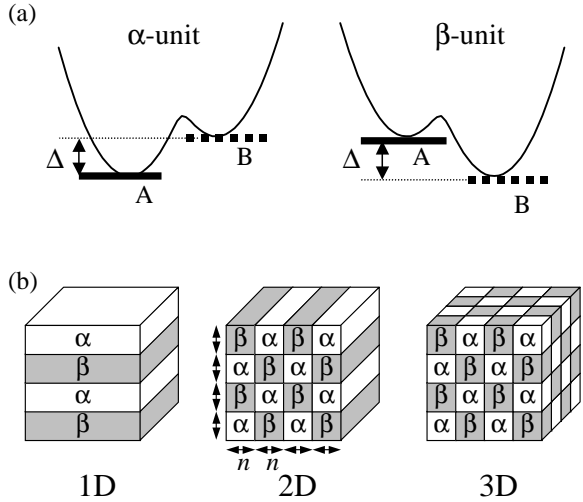}
\end{center}
\caption{
(a) Schematic bistable potentials with opposite relative stabilities in $\alpha$ and $\beta$ units. 
(b) Schematic one-, two-, three-dimensional superlattice structures consisting of the $\alpha$ and $\beta$ blocks. }
\label{fig:PotentialAndSuperlattice}
\end{figure}

We consider a phenomenological model described by the Ising Hamiltonian\cite{Boukheddaden00PRB,Enachescu01JP}
\begin{equation}
H = -\sum_{(i,j)={\rm n.n.}} J_{ij}S_iS_j 
  + \sum_{i}\varepsilon_iS_i,	\label{eq:Hamiltonian}
\end{equation}
on a three-dimensional lattice, 
where the `spin' variable $S_i=\pm 1$ denotes the two states (B and A) of the site $i$ with the energy difference $2\varepsilon_i$. 
The nearest neighbor (n.n.) coupling $J_{ij}$ is assumed to be a positive constant $J$, so that the neighboring sites prefer to be in the same state. 
We use a simple cubic lattice of $N \times N \times N$ sites with periodic boundary conditions. 
$N=60$ is a typical size used in our simulations. 
We consider the superstructures consisting of two kinds of blocks, $\alpha$ and $\beta$, alternating in one-, two-, or three-dimensional (1D, 2D, 3D) directions as shown in Fig.~\ref{fig:PotentialAndSuperlattice}(b). 
The size of each block is $n$ in the confined directions. 
(The system size $N$ must be an integer multiple of $2n$.)
The site potential $\varepsilon_i$ is assumed to be constant in each of the blocks, $-\Delta/2$ and $\Delta/2$ in the $\alpha$ and $\beta$ blocks, respectively, corresponding to the situation illustrated in Fig.~\ref{fig:PotentialAndSuperlattice}(a). 
Note that the system is symmetric with respect to the interchange of A and B ($\alpha$ and $\beta$). 
The case of $\Delta=0$ corresponds to the uniform system. 

For $\Delta>0$, the $\alpha$ and $\beta$ blocks favor the A and B states, respectively. 
However, the system still tends to be in a uniform state without phase separation, if the block thickness $n$ is not very large and the {\it inter}-block interface energy proportional to $J$ outweighs the {\it intra}-block energy gain proportional to $|\Delta|$.  
The condition for uniformity is written as $|\Delta| < \Delta_{\rm c} \equiv 4dJ/n$, where $d$ is the dimensionality of the superstructure. 
Under this condition, the uniform ground states are doubly degenerate: The phase of all the sites being in the A state has the same energy as that of all the sites in the B state. 

We have studied the kinetics of the model at finite temperatures under external excitations by means of a classical Monte Carlo simulation\cite{Binder79Book}, wherein the spin flipping rate at each site $j$ is the sum of the thermal and external terms: $w_j=w_j^{\rm T}+w^{\rm P}_j$. 
When we consider selective excitation from A to B, we set $w^{\rm P}_j(A\rightarrow B)=W$ and $w^{\rm P}_j(B\rightarrow A)=0$. 
In each Monte Carlo step (MCS), both thermal and external spin-flip trials are performed at every site once on average. 
The heat bath model is used for the thermal flipping rate $w_j^{\rm T}$, 
although we have checked that other algorithms such as the Glauber, Metropolis, and Arrhenius models give essentially the same results. 

\begin{figure}
\begin{center}
  \leavevmode
  \epsfxsize=85mm
  \epsfbox{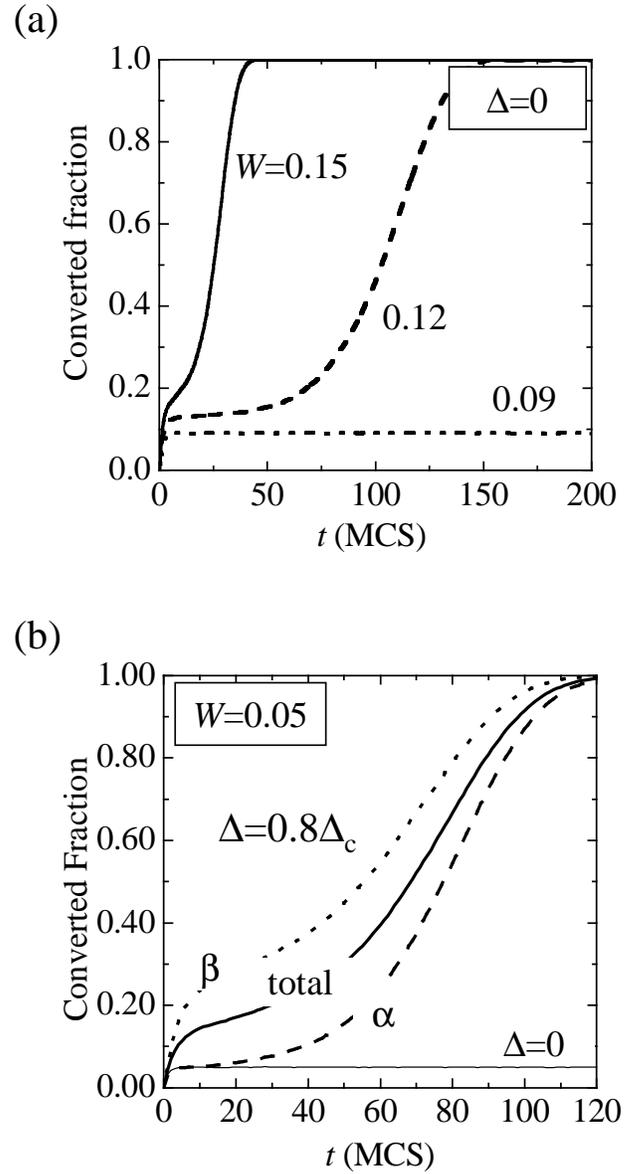}
\end{center}
\caption{
Monte Carlo simulation on the kinetics of the excitation-induced phase transition in the system of $60 \times 60 \times 60$ sites at $k_{\rm B}T=J$. 
(a) The fraction of converted sites as a function of the excitation time $t$ in Monte Carlo steps (MCS) with $\Delta = 0$ for various excitation rate $W$ per MCS. 
(b) The converted fraction in the 2D superstructure with $n$ = 1 for $W=0.05$ per MCS at $k_{\rm B}T=J$, in the case of $\Delta=0.8\Delta_{\rm c}=6.4J$ (thick line) compared with the case of $\Delta=0$ (thin line).  
Solid, broken and dotted thick lines show the total fraction and the partial fractions of the $\alpha$ and $\beta$ blocks, respectively.  
}
\label{fig:Fraction_on_Time}
\end{figure}

Suppose the system is initially in the A phase and develops under an excitation rate $W$ from A to B at temperature $k_{\rm B}T=J$. 
Figure~\ref{fig:Fraction_on_Time}(a) shows the fraction of sites in the B state as a function of the excitation time $t$ for various $W$ in the homogeneous case of $\Delta=0$. 
For weak excitation with $W=0.09$, the system remains permanently in the A phase, whereas conversion to the B phase occurs for larger $W$. 
The converted fraction does not increase linearly with time but grows rather suddenly after a certain time. 
This is a clear reflection of the cooperative nature of the phase change. 
This type of behavior was actually observed in a spin-crossover complex, where the transition occurs with the absorbed photon flux of about 10$^{18}$ cm$^{-3}$s$^{-1}$\cite{Ogawa00PRL}. 
  
We move on to the case of a superstructure with $\Delta\ne 0$. 
Figure~\ref{fig:Fraction_on_Time}(b) displays the time evolution for a weak excitation rate $W=0.05$ at $k_{\rm B}T=J$ in the 2D superstructure with the block size $n=1$ and $\Delta=0.8\Delta_{\rm c}=6.4J$. 
The phase change occurs in a short time in contrast to the case of $\Delta=0$, for which there is no indication of a phase transition. 
(Note the different time scales of Fig.2(a) and Fig.2(b).)

To clarify the underlying mechanism of the accelerated phase transition, we plot in Fig.~\ref{fig:Fraction_on_Time}(b) the converted fractions in the $\alpha$ blocks and in the $\beta$ blocks separately.
The converted fraction of the $\beta$ blocks grows quickly from the beginning, and then that of the $\alpha$ blocks gradually follows. 
By tracking the time evolution of all the sites, it has been found that a nucleated domain of B states in a $\beta$ block does not decay so quickly as in the case of the uniform structure because of the local stabilization energy $\Delta$, thereby allowing further growth of the B domain around the nucleus firstly in the $\beta$ block and then onto the surrounding $\alpha$ blocks. 
Although the energy difference $\Delta$ counteracts as an energy cost in the $\alpha$ blocks, the growth process continues once the stable nucleus is formed. 

\begin{figure}
\begin{center}
  \leavevmode
  \epsfxsize=85mm
  \epsfbox{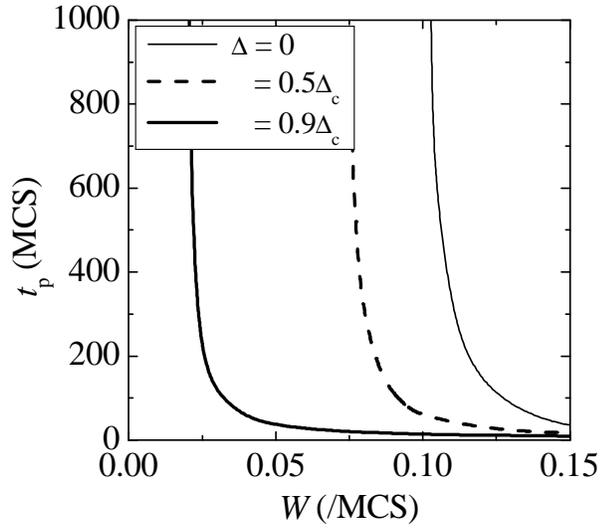}
\end{center}
\caption{
Dependence of the conversion time $t_{\rm p}$ on the excitation intensity $W$ in the 2D superstructure with $n=1$ at $k_{\rm B}T=J$ for various $\Delta$ with $\Delta_{\rm c}=8J$. }
\label{fig:tp_on_W}
\end{figure}

The time of continuous excitation required for inducing the phase transition, $t_{\rm p}$, depends on the excitation intensity $W$. 
The $W$-dependence of $t_{\rm p}$ is shown in Fig.~\ref{fig:tp_on_W} for various $\Delta$ in the 2D superstructure. 
Here $t_{\rm p}$ is defined by the time of the converted fraction exceeding 0.8. 
There is a threshold $W_{\rm c}$ below which the phase transition does not occur for each $\Delta$, and $t_{\rm p}$ sharply diverges from above $W_{\rm c}$. 
We have confirmed the critical behavior by extending simulations up to 100,000 MCSs. 
Similar results have been obtained for 1D and 3D superstructures. 
The existence of a threshold light intensity for the photoinduced phase transition was also suggested in real materials\cite{Ogawa00PRL,Suzuki99PRB}.
The threshold $W_{\rm c}$ at which $t_{\rm p}$ diverges is reduced drastically with increasing $|\Delta|$. 
The results in Fig.~\ref{fig:tp_on_W} imply that the required time $t_{\rm p}$ at each $W$ drastically decreases with increasing $|\Delta|$. 
For example, at $W$=0.12, the phase transition in the case of $\Delta=0.9\Delta_{\rm c}$ requires only one tenth of the excitation time for the homogeneous structure with the same excitation intensity. 

In conclusion, by using nanostructures, we can drastically reduce the threshold light intensity and enhance the efficiency of phase switching in both the forward and backward directions. 
With this principle, which does not rely on the details of materials, the range of useful materials can be extended substantially, because {\it an appropriate combination of inferior materials can produce a superior material} for photoinduced switching.

The present mechanism was hinted by the study of the photoinduced magnetism in cobalt-iron cyanides \cite{Sato96Science2,Sato97JES,Goujon00EPJB,Bleuzen00JACS,Pejakovic00PRL,Kawamoto01PRL}, which can be considered as a mixed system of different local environments with cooperative interactions due to lattice distortions\cite{Kawamoto01PRL}. 
It would be very interesting to make a superstructure in this class of materials as well as in other materials such as perovskite manganese oxides\cite{Sun96APL,Lu00PRB}, for which a photoinduced metal-insulator transition was observed in the bulk\cite{Miyano97PRL}.

The authors acknowledge useful discussions with Kazuhito Hashimoto, Naonobu Shimamoto, Osamu Sato and Yoshihiro Asai. 
This work was partly supported by NEDO under the Nanotechnology Materials Program.
Computations were carried out partly using the facilities of the AIST Tsukuba Advanced Computing Center.



\end{document}